\documentclass[final,authoryear,5p,times,twocolumn]{elsarticle}
\usepackage{amsmath}
\usepackage{amssymb} 

\usepackage{tikz}
\usetikzlibrary{arrows,matrix,positioning,patterns}
\usetikzlibrary{calc}
\usepackage{pgfplots} 
\usetikzlibrary{external}
\usetikzlibrary{plotmarks}

\newdefinition{rmk}{Remark}
\newproof{pf}{Proof}
\newproof{pot}{Proof of Theorem \ref{thm2}}

\usepackage{algorithm} 
\usepackage{algpseudocode}

\usepackage{amsthm}

\usepackage{amsmath}
\usepackage{amssymb} 

\usepackage{eurosym}
\usepackage{colortbl}

\usetikzlibrary{arrows.meta,calc,decorations.markings,math,arrows.meta}

\usepackage[labelformat=simple]{subcaption}
\DeclareCaptionLabelSeparator{periodspace}{.\quad}
\usepackage{color,framed}
\usepackage[utf8]{inputenc} 
\usepackage{tabularx}

\usepackage{rotating}
\usepackage{graphicx}
\usepackage{lscape}
\allowdisplaybreaks 
\usepackage{longtable}

\usepackage[utf8]{inputenc}
\usepackage{pgfplots}
\usepgfplotslibrary{groupplots} 
\pgfplotsset{compat=1.3}

\makeatletter
\def\ps@pprintTitle{%
 \let\@oddhead\@empty
 \let\@evenhead\@empty
 \def\@oddfoot{}%
 \let\@evenfoot\@oddfoot}
\makeatother

\begin{document}

\begin{frontmatter}
\title{Optimal In-field Routing for Full and Partial Field Coverage\\ with Arbitrary Non-Convex Fields and Multiple Obstacle Areas}

\author{Mogens Graf Plessen\corref{cor1}}
%
%
\cortext[cor1]{\texttt{mgplessen@gmail.com}}
%

\begin{abstract}
Within the context of optimising the logistics in agriculture this paper relates to optimal in-field routing for full and partial field coverage with arbitrary non-convex fields and multiple obstacle areas. It is distinguished between nine different in-field routing tasks: two for full-field coverage, seven for partial-field coverage and one for shortest path planning between any two vertices of the transition graph. It differentiates between equal or different start and end vertices for a task, coverage of only a subset of vertices, and a subset of edges or combinations. The proposed methods are developed primarily for applying sprays and fertilisers with larger operating widths and with fields where there is unique headland path. Partial field coverage where, e.g., only a specific subset of edges has to be covered is relevant for precision agriculture and also for optimised logistical operation of smaller-sized machinery with limited loading capacities. The result of this research is the proposition of two compatible algorithms for optimal full and partial field coverage path planning, respectively. These are evaluated on three real-world fields to demonstrate their characteristics and computational efficiency.
\end{abstract} 
\end{frontmatter}


\section{Introduction and Problem Formulation\label{sec_intro}}

\subsection{Motivation\label{subsec_motivation}}

Within the context of optimising logistics in agriculture this paper addresses the research question: how to optimally solve multiple in-field routing tasks including both \emph{full} and \emph{partial} field coverage, while simultaneously accounting for non-convex fields, multiple obstacle areas, partitioned subfields, and compacted area minimisation? It is thus aimed at optimising in-field coverage path planning, while generalising this to different tasks and different field characteristics.

\subsection{Problem Formulation and Contribution\label{subsec_problFormul}}

The problem addressed is to solve the following nine in-field routing tasks: (i) full field coverage with equal start and end vertex, (ii) full field coverage with different start and end vertex, (iii) coverage of a subset of vertices with equal start and end vertex, (iv) coverage of a subset of vertices with different start and end vertex, (v) coverage of a subset of edges with equal start and end vertex, (vi) coverage of a subset of edges with different start and end vertex, (vii) a combination of (iii) and (v), (viii) a combination of (iv) and (vi), and (ix) shortest path planning between any two vertices. As a starting point, the following assumptions are made: availability of a transition graph with edges connecting vertices, a corresponding cost array with edge-costs equal to the path length of edges, a designated start and end vertex, implementation of the task by one vehicle (instead of multiple in-field operating vehicles), and any of above  in-field routing tasks (i)-(ix). 

For this setup, the contribution of this paper is to address the research question from section \ref{subsec_motivation}.

\begin{figure}
\centering
\begin{tikzpicture}
\draw[dashed] (0.25, -0.09) -- (0.25, 2.66);
\draw[draw=black,fill=black] (0.25, -0.09) circle (0.75pt); 
\draw[draw=black,fill=black] (0.25, 2.66) circle (0.75pt); 
\node[color=black] (a) at (0.25, -0.33) {\small{$1$}};
\draw[dashed] (0.55, -0.1) -- (0.55, 2.7);
\draw[draw=black,fill=black] (0.55, -0.18) circle (0.75pt); 
\draw[draw=black,fill=black] (0.55, 2.76) circle (0.75pt); 
\node[color=black] (a) at (0.55, -0.39) {\small{$2$}};
\draw[dashed] (0.85, -0.19) -- (0.85, 2.79);
\draw[draw=black,fill=black] (0.85, -0.28) circle (0.75pt); 
\draw[draw=black,fill=black] (0.85, 2.79) circle (0.75pt);
\draw[dashed] (1.15, -0.29) -- (1.15, 2.82);
\draw[draw=black,fill=black] (1.15, -0.37) circle (0.75pt); 
\draw[draw=black,fill=black] (1.15, 2.825) circle (0.75pt); 
\draw[dashed] (1.45, -0.32) -- (1.45, 2.86);
\draw[draw=black,fill=black] (1.45, -0.385) circle (0.75pt); 
\draw[draw=black,fill=black] (1.45, 2.86) circle (0.75pt); 
\draw[dashed] (1.75, -0.23) -- (1.75, 2.9);
\draw[draw=black,fill=black] (1.75, -0.335) circle (0.75pt); 
\draw[draw=black,fill=black] (1.75, 2.89) circle (0.75pt); 
\draw[dashed] (2.05, -0.12) -- (2.05, 2.96);
\draw[draw=black,fill=black] (2.05, -0.19) circle (0.75pt); 
\draw[draw=black,fill=black] (2.05, 2.92) circle (0.75pt); 
\node[color=black] (a) at (2.22, -0.39) {\small{...}}; 
\draw[dashed] (2.35, -0.02) -- (2.35, 2.95);
\draw[draw=black,fill=black] (2.35, -0.02) circle (0.75pt); 
\draw[draw=black,fill=black] (2.35, 2.945) circle (0.75pt); 
\draw[draw=black,fill=black] (0.04, 1.3) circle (0.75pt);
\node[color=black] (a) at (-0.12, 1.3) {\small{$0$}};  
\draw[draw=black,fill=black] (2.553, 1.6) circle (0.75pt);
\draw[solid,rounded corners=0.5cm] (0, 0) -- (1.5, -0.5) -- (2.75,0.2)--(2.5,1.5)--(2.75,3)--(0,2.7)--(0.1,1.5)--(-0.1,0.6)--(0,0);
\draw[dashed] (3.8, -0.03) -- (3.8, 0.81);
\draw[draw=black,fill=black] (3.8, -0.065) circle (0.75pt); 
\draw[draw=black,fill=black] (3.8, 0.78) circle (0.75pt); 
\draw[dashed] (4.1, -0.09) -- (4.1, 0.98);
\draw[draw=black,fill=black] (4.1, -0.165) circle (0.75pt); 
\draw[draw=black,fill=black] (4.1, 1.01) circle (0.75pt); 
\draw[dashed] (4.4, -0.17) -- (4.4, 1.17);
\draw[draw=black,fill=black] (4.4, -0.265) circle (0.75pt); 
\draw[draw=black,fill=black] (4.4, 1.22) circle (0.75pt); 
\draw[dashed] (4.1, 2.34) -- (4.1, 3.1);
\draw[draw=black,fill=black] (4.1, 2.32) circle (0.75pt); 
\draw[draw=black,fill=black] (4.1, 3.12) circle (0.75pt); 
\draw[dashed] (4.4, 2.15) -- (4.4, 3.2);
\draw[draw=black,fill=black] (4.4, 1.97) circle (0.75pt); 
\draw[draw=black,fill=black] (4.4, 3.2) circle (0.75pt); 
\draw[dashed] (4.7, -0.3) -- (4.7, 3.2);
\draw[draw=black,fill=black] (4.7, -0.36) circle (0.75pt); 
\draw[draw=black,fill=black] (4.7, 3.23) circle (0.75pt); 
\draw[dashed] (5, -0.35) -- (5, 3.2);
\draw[draw=black,fill=black] (5, -0.383) circle (0.75pt); 
\draw[draw=black,fill=black] (5, 3.22) circle (0.75pt); 
\draw[dashed] (5.3, -0.35) -- (5.3, 3.17);
\draw[draw=black,fill=black] (5.3, -0.34) circle (0.75pt); 
\draw[draw=black,fill=black] (5.3, 3.165) circle (0.75pt); 
\draw[dashed] (5.6, -0.2) -- (5.6, 3.1);
\draw[draw=black,fill=black] (5.6, -0.203) circle (0.75pt); 
\draw[draw=black,fill=black] (5.6, 3.11) circle (0.75pt); 
\draw[dashed] (5.9, -0.02) -- (5.9, 3);
\draw[draw=black,fill=black] (5.9, -0.02) circle (0.75pt); 
\draw[draw=black,fill=black] (5.9, 3.05) circle (0.75pt);
\draw[draw=black,fill=black] (3.825, 2.75) circle (0.75pt); 
\draw[draw=black,fill=black] (3.58, 0.3) circle (0.75pt); 
\draw[draw=black,fill=black] (6.17, 2.65) circle (0.75pt); 
\draw[solid,rounded corners=0.5cm] (3.6, 0) -- (5.1, -0.5) -- (6.28,0.2)--(5.9,1.5)--(6.28,3)--(4.5,3.3)--(3.6,2.9)--(4.8,1.5)--(3.5,0.6)--(3.6,0);
\draw[solid,rounded corners=0.1cm,fill=white] (4.9, 2.45) -- (5.2, 2.4) --  (5.43, 2.45) -- (5.43,2.65)--(5.2, 2.7) -- (4.9,2.65)--(4.9,2.45);
\draw[solid,rounded corners=0.1cm,fill=white] (4.4,0.25)--(5.15,0.45)--(5.05,0.7)--(4.2,0.4)--(4.4,0.25); 
\draw[draw=black,fill=black] (4.398, 0.473) circle (0.75pt); 
\draw[draw=black,fill=black] (4.398, 0.26) circle (0.75pt); 
\draw[draw=black,fill=black] (4.7, 0.578) circle (0.75pt);
\draw[draw=black,fill=black] (4.7, 0.33) circle (0.75pt);
\draw[draw=black,fill=black] (5, 0.675) circle (0.75pt); 
\draw[draw=black,fill=black] (5, 0.41) circle (0.75pt); 
\draw[draw=black,fill=black] (5, 2.667) circle (0.75pt);
\draw[draw=black,fill=black] (5, 2.435) circle (0.75pt);
\draw[draw=black,fill=black] (5.3, 2.68) circle (0.75pt);
\draw[draw=black,fill=black] (5.3, 2.42) circle (0.75pt); 
%
\end{tikzpicture}
\caption{Illustration of different types of field shapes and the notion of tractor tracks defining a transition graph $\mathcal{G}$. Headland and interior edges are denoted by solid and dashed lines, respectively. Vertices are indicated by dots and are, in general, labelled for identification such that the transition between any two vertices is unique. (Left) Uninterrupted edges when aligned in a rotated coordinate frame. (Right) Interruped edges due to field indents and obstacle areas that are prohibited from tresspassing by any vehicle operating in the field.}
\label{fig_2FieldShapes}
\vspace{-0.5cm}
\end{figure}
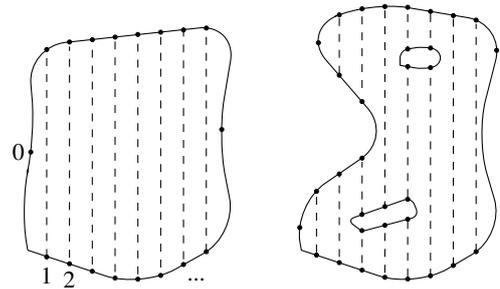

\subsection{Background and Further Motivation}

According to \cite{ahumada2009application} there are four main functional sectors for the agri-food supply chain: production, harvesting, storage and distribution. Optimising logistics and route planning plays an important role  in all of the four functional areas for improved supply chain efficiency. Furthermore, according to \cite{sorensen2010conceptual} it can be distinguished between in-field, inter-field, inter-sector and inter-regional logistics. This paper relates to the first functional area of the agri-food supply chain, i.e., production, and further to in-field optimisation of logistics. 

The classic vehicle routing problem (VRP) seeks total cost-minimising routes for multiple identical vehicles that all start and end at a single depot and are subject to load constraints, and where multiple vertices (customers) subject to various demands must be serviced exactly once by exactly one vehicle. There are many variations, see \cite{toth2014vehicle}. The focus is on \emph{vertex}-coverage. By contrast, for agricultural in-field routing typically \emph{edge}-coverage is of primary importance. Often a single large machine is operating in-field, e.g., during a spraying application. Therefore, instead of VRPs, arc routing problems (ARPs) are here of more interest, see \cite{eiselt1995arc,eiselt1995arc2}.  

For shortest path planning between two vertices (routing task (ix) according to sect. \ref{subsec_problFormul}) greedy
algorithms such as the algorithm by \cite{dijkstra1959note} and the A$^*$-algorithm by \cite{hart1968formal} are famous. However, these are specialised algorithms that do not solve field coverage path planning problems where a set of edges needs to be covered. For ARPs according to \cite{eiselt1995arc,eiselt1995arc2}, problems can be distinguished between the classes of the Chinese postman problem (CPP) and the rural postman problem (RPP), where all and only a subset of all arcs of the graph need to be traversed, respectively. For the CPP it is further distinguished between undirected, directed, windy, mixed and hierarchical CPPs. Similarly, this occurs for RPPs. To further illustrate complexity of solution algorithms and their typical hierarchical structure, a directed RPP can be solved by first constructing a shortest spanning arborescence, then deriving an Eulerian graph on top, and ultimately determining an Eulerian tour on the augmented graph. In general, Eulerian tours on an Eulerian graph are not unique and can differ substantially. Therefore, specific heuristics can be derived to develop path planning behaviour.

Given an agricultural working area, the first step is to fit a transition graph with edges and vertices as their connection points. This step requires to (i) decide on field-interior edge shapes, straight or curved, and (ii) to account for field contours and possibly also for 3D topography, see \cite{hameed2013optimized}. According to sect.~\ref{subsec_problFormul}, this paper assumes a transition graph as a given starting point, whereby obstacle areas prohibited from trespassing are also accounted for according to Fig. \ref{fig_2FieldShapes}. 

\cite{griffel2018machinery} and \cite{seyyedhasani2019routing} examined the role of field shapes on efficiency of path planning. \cite{zhou2014agricultural} discussed a hierarchical and heuristic algorithm for in-field routing with obstacle areas. It was hierarchical since it decomposed the problem into three sequential stages. It was heuristic since it decomposed the field area firstly into cells, then determined a sequence for coverage of the cells, and only then considered path plans and their linking between the different cells. A similar method was described in \cite{taix2003path}. Importantly, because of these hierarchical heuristics and for transition graphs applied to arbitrary field shapes and arbitrarily located multiple obstacles, these methods can in general not guarantee to find a minimum cost tour covering all edges at least once.  This is mentioned to explicitly  stress that, by contrast, the algorithm proposed in this paper is guaranteed to find the minimum cost tour. This is achieved by working directly on the full Eulerian graph augmentation. Heuristic rules are derived on top to guide the planning of an Eulerian tour with favourable properties for additional partial field coverage and practical implementation.


\begin{table}
\centering
\begin{tabular}{|ll|}
\hline
\multicolumn{2}{|c|}{MAIN NOMENCLATURE}\\
$\mathcal{E}$ & Set of all edges of $\mathcal{G}$: $(i,j)\in\mathcal{E}$. \\
$\mathcal{E}'$ & Replicated edges added to $\mathcal{G}$ to produce $\mathcal{G}'$. \\
$\mathcal{E}_{\text{hdl},\text{hdl}}^\text{hdl}$ & Set of headland edges with vertices $i,j\in\mathcal{V}^\text{hdl}$. \\
$\mathcal{E}_{\text{isl},\text{isl}}^\text{hdl}$ & Set of island headland edges with $i,j\in\mathcal{V}^\text{isl}$. \\
$\mathcal{E}_{\text{hdl},\text{isl}}^\text{int}$ & Set of interior edges with $i\in\mathcal{V}^\text{hdl}$, $j\in\mathcal{V}^\text{isl}$. \\
$\mathcal{E}_{\text{isl},\text{isl}}^\text{int}$ & Set of interior edges with $i\in\mathcal{V}^\text{isl}$, $j\in\mathcal{V}^\text{isl}$. \\
$\mathcal{G}$ & Undirected transition graph, $\mathcal{G}=(\mathcal{V},\mathcal{E})$. \\
$\mathcal{G}'$ & Eulerian graph augmentation of $\mathcal{G}$. \\
$\mathcal{I}$ & List of indices  for partial field coverage.\\
$\mathcal{L}_\mathcal{V}$ & Subset of vertices for partial field coverage.\\
$\mathcal{L}_\mathcal{E}$ & Subset of edges for partial field coverage.\\
$\mathcal{P}$ & Ordered list relevant for partial field coverage.\\
$\mathcal{T}^\text{abu}$ & Tabu list of instances of $\mathcal{I}$, relevant in Algorithm \ref{alg_mainAlg_T3toT8}.\\
$\mathcal{V}$ & Set of all vertices of $\mathcal{G}$. \\
$\mathcal{V}^\text{hdl}$ & Set of headland (hdl) vertices of $\mathcal{G}$. \\
$\mathcal{V}^\text{isl}$ & Set of island (isl) vertices of $\mathcal{G}$. \\
$C$ & Accumulated cost of a vertex-sequence, (m).\\
$\Delta$AB & Path length savings w.r.t. AB-pattern, (m, $\%$).\\
$N_\mathcal{I}$, $N_{\mathcal{T}^\text{abu}}$ & 2 scalar hyperparameters in Algorithm \ref{alg_mainAlg_T3toT8}.\\
$c_{i,j}$ & Edge-cost (path length) for $(i,j)\in\mathcal{E}$, (m).\\
$s_t$ & Vertex of a sequence $\{s_t\}_{0}^T$ at index $t$.\\
$\{s_t\}_{0}^T$ & Sequence of vertices for $t=0,\dots,T$.\\
$\{s_t^\text{pfc}\}_{0}^{T^\text{pfc}}$ & Sequence of vertices for partial field coverage.\\
$\{s_t^\text{sub}\}_{0}^{T^\text{sub}}$ & Subsequence of vertices for $t=0,\dots,T^\text{sub}$.\\
$s_\text{start},s_\text{end}$ & Start and end vertex for a routing task.\\
\hline
\end{tabular}
\vspace{-0.5cm}
\end{table}

Field decomposition into subfields using trapezoids as presented in \cite{oksanen2009coverage} is a popular method to deal with irregularly shaped fields (\cite{mederle2017analysis} and \cite{santoro2017route}). 
 
\cite{bochtis2008minimising}, \cite{bochtis2013benefits} and \cite{zhou2015quantifying}, discussed different fieldwork patterns and headland turning methods subject to vehicle kinematic constraints. They were typically motivated by the desire to minimise accumulated non-working path lengths at headlands. Note that these methods do not naturally account for in-field obstacles. Instead, as mentioned in the introduction of \cite{bochtis2013benefits}, \textit{``B-patterns do not generate any subfield areas division but, by contrast, the generation of the subfields (when it is needed due to e.g. physical obstacles or complex field shapes) is a prerequisite for applying B-patterns methodology''}. As a consequence, when accounting for obstacles (such as tree islands), B-patterns are in general no longer optimal since the same limitations of aforementioned hierarchical algorithms apply.   

For a discussion of field experiments over three years for different headland turning methods see \cite{paraforos2018automatic}.

In \cite{plessen2018partial} two path planning patterns for partial field coverage were compared for convex field shapes. One of them was identified as particularly favourable when aiming for minimal compacted area from tractor tracks while accounting for limited turning radii of agricultural machinery. However, patterns are in general never optimal for arbitrary non-convex field shapes particularly when also considering multiple obstacle areas. Thus, the present paper is generalising and relevant not only for full but also for partial field coverage. Furthermore, as will be shown, algorithms are designed purposely such that the preferred pattern from \cite{plessen2018partial} is automatically recovered for convex field shapes.

The remaining paper is organised as follows: algorithms, real-world examples, benefits and limitations, and the conclusion are described in sections \ref{sec_solnDescription}-\ref{sec_conclusion}.

\section{Solution Description\label{sec_solnDescription}}

\subsection{High-level strategy to address nine different routing tasks}

The nine classes of different in-field routing tasks addressed in this paper are summarised in Table \ref{tab_9routingTasks}. In practice, this number is required for generality. For perspective, spraying applications occur multiple times throughout any crop year. Depending on available machinery, weather and varying available time-windows, different routing tasks may apply, also including partial field coverage per field run. In view of precision agriculture, algorithmic solutions are therefore needed to address all of tasks, T1 to T9.

In this paper, two main levels are distinguished. At the highest level, there are the \emph{full} field coverage tasks T1 and T2, whereby T2 can be considered as an extension of T1. The algorithm proposed therefore is discussed in sect. \ref{subsec_T1andT2}. On the second level, there are the \emph{partial} field coverage tasks T3 to T8. Here, tasks T3, T5 and T7, and equivalently tasks T4, T6 and T8, exploit as their starting points the full field coverage solutions for T1 and T2, respectively. For the second level, the proposed algorithm is discussed in sect. \ref{subsec_T3toT8}. Finally, T9 is a special case that is discussed in sect. \ref{subsec_T9}.

\begin{table}
\centering
\begin{tabular}{|ll|}
\hline
\multicolumn{2}{|c|}{9 ROUTING TASKS}\\
T1 & Full field coverage with $s_\text{start}==s_\text{end}$.\\
T2 & Full field coverage with $s_\text{start}\neq s_\text{end}$.\\
T3 & Coverage of a subset of vertices with $s_\text{start}==s_\text{end}$.\\
T4 & Coverage of a subset of vertices with $s_\text{start}\neq s_\text{end}$.\\
T5 &  Coverage of a subset of edges  with $s_\text{start}==s_\text{end}$.\\
T6 &  Coverage of a subset of edges with $s_\text{start}\neq s_\text{end}$.\\ 
T7 & A combination of T3 and T5.\\ 
T8 & A combination of T4 and T6.\\
T9 & Shortest path planning between any 2 vertices of $\mathcal{G}$.\\ 
\hline
\end{tabular}
\caption{In this paper, 9 classes of in-field routing tasks are considered.}
\label{tab_9routingTasks}
\end{table}

\begin{figure}
\centering
\includegraphics[width=8.5cm]{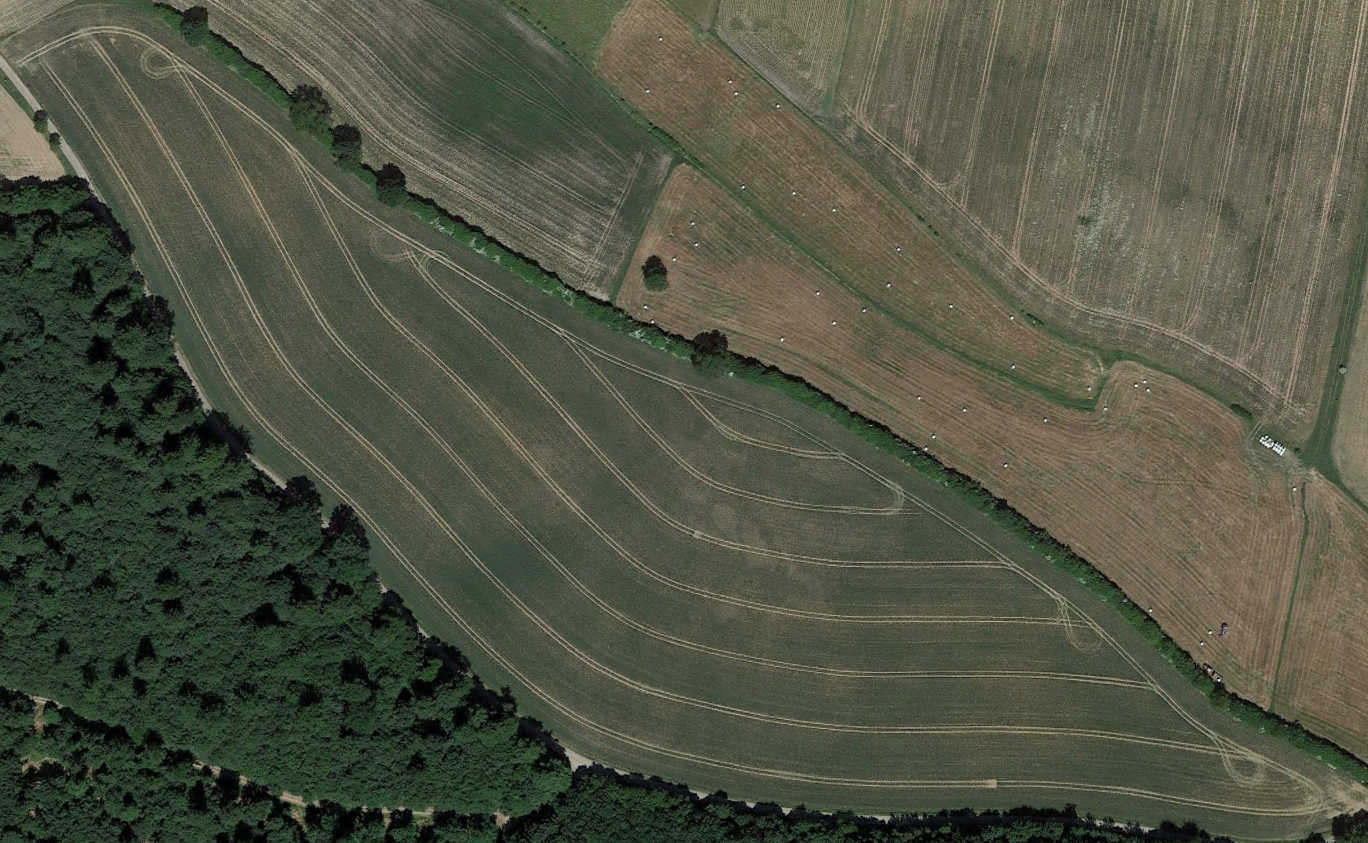} 
\caption{Illustration of curved interior edges aligned to part of the field contour.}
\label{fig_curved}
\end{figure}

\subsection{Preliminaries\label{subsec_preliminaries}}

\begin{figure}
\centering
\includegraphics[width=8.7cm]{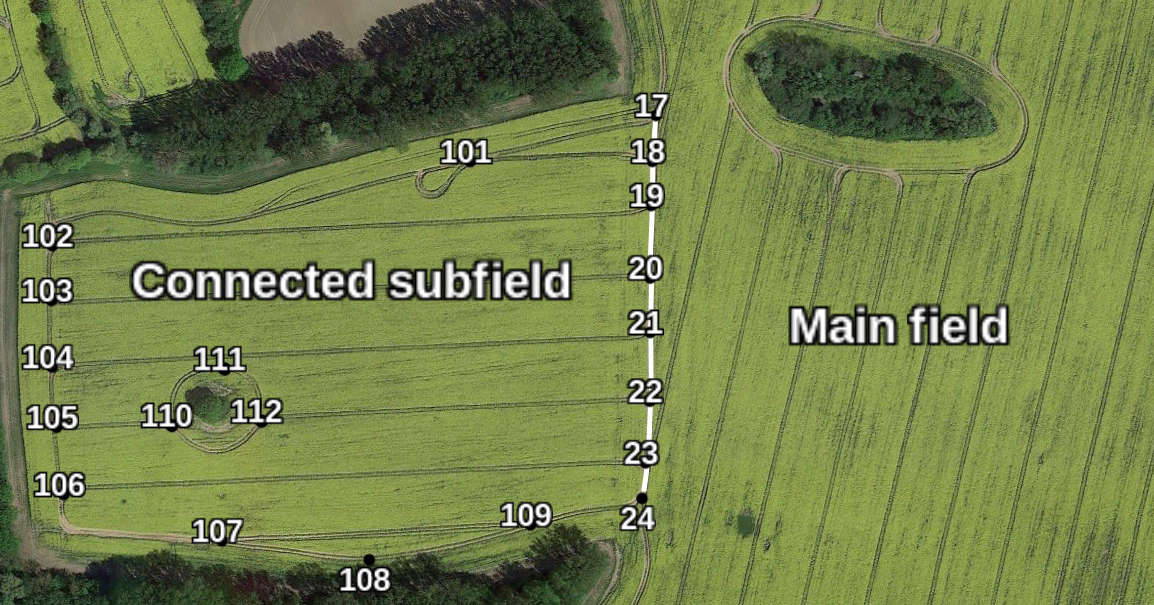} 
\caption{Illustration of a subfield connected to the main field. Characteristic is (i) the sharing of a path, here from vertex 17 to 24, that is coinciding for both subfield and main field, and (ii) different interior edge orientations. The partition into main field and subfields occurs in particular in case of strongly non-convex fields, where it is worthwhile to differentiate interior edge orientations to minimise compacted area. The annotation is added for two reasons: (i) to illustrate vertex labelling on a satellite image, (ii) to emphasise that connected subfields and main field are treated separately according to sect. \ref{subsec_connectedSubfields}.}
\label{fig_subfield}
\vspace{-0.3cm}
\end{figure}

Topographical characteristics relevant for in-field routing, data variables, and solution variables are discussed. For the former, these are (i) arbitrarily shaped fields (convex or non-convex), (ii) multiple obstacle areas\footnote{The term  ``obstacle area'' comprises all obstacles prohibited from trespassing by in-field operating vehicles, including tree islands, ponds and so forth.} within the field, (iii) either straight or curved interior edges aligned to part of the field contour, and (iv) partitioned subfields with interior edges orientated differently from the area-wise largest main part of the field. These are shown in Figs. \ref{fig_curved} and \ref{fig_subfield}. Two more comments should be made. Firstly, curved interior edges still permit a transition graph representation that is analogous to Fig. \ref{fig_2FieldShapes}. Secondly, in this paper separate transition graphs are defined for all partitioned subfields and the main field. The synchronised handling of all of these is detailed in sect. \ref{subsec_connectedSubfields}. 

Data variables used for the problem description are summarised in the main nomenclature table. See also Fig. \ref{fig_2FieldShapes} for visualisation. Some additional comments are made. First, for every undirected transition graph, $\mathcal{G}=(\mathcal{V},\mathcal{E})$, the set of vertices and edges can be partitioned into subsets as follows:
\begin{subequations}
\begin{align}
\mathcal{V} &= \mathcal{V}^\text{hdl} \cup \mathcal{V}^\text{isl},\\
\mathcal{E} &= \mathcal{E}_{\text{hdl},\text{hdl}}^\text{hdl} \cup \mathcal{E}_{\text{isl},\text{isl}}^\text{hdl} \cup \mathcal{E}_{\text{hdl},\text{isl}}^\text{int} \cup \mathcal{E}_{\text{isl},\text{isl}}^\text{int}.
\end{align}
\end{subequations}
Second, start and end vertex, $s_\text{start}$ and $s_\text{end}$, typically denote the field entry and exit vertex. Third, a list of elements is denoted by $\{\cdot \}$, the number of elements in a list by $|\cdot|$, and an edge between vertices $i$ and $j$ by $(i,j)$. The ``$+$''-operator is overloaded to indicate concatenation of lists as $\{\cdot \}+\{\cdot \}$. Fourth, throughout this paper, it is assumed that only \emph{forward motion} of any in-field operating machinery is permitted. Thus, any sequence of vertices such as \textit{a-b-a} is prohibited. For shortest path planning it would imply the necessity of reverse driving, which is impractical, in particular for operations with trailers and for large edge lengths. Fifth, as Fig. \ref{fig_2FieldShapes} illustrates, the number of edges incident to every vertex is either two or three. However, because  of the previous assumption about forward motion only, at every vertex there are always only either one or two transition decisions available during routing. Ultimately, there always exists a path length minimising global optimal solution to all routing tasks T1 to T9. This immediately follows from the nonnegativity property of edge-costs,  here defined as path lengths.

Solution variables most relevant in the presented algorithms are $s_t$, $\{s_t\}_{0}^{T}$, $C$, and variations with different superscripts. These indicate a vertex at index $t$, a sequence of vertices, and the accumulated path length, respectively.

\subsection{Full Field Coverage: T1 and T2\label{subsec_T1andT2}}

Algorithm \ref{alg_mainAlg_T1T2} is proposed for full field coverage path planning according to routing tasks T1 and T2.

\begin{algorithm}
\caption{Full Field Coverage for T1 and T2}
\begin{tabular}{ l l l l l}
0. & \multicolumn{4}{l}{\textbf{Required Subfunctions} (brief description): }\\[0.1cm]
  & \multicolumn{4}{l}{$\mathcal{F}^\text{hdl}(\cdot)$: for tracing of the headland path.}\\[0.1cm]
  & \multicolumn{4}{l}{$\mathcal{F}^\text{fsp}(\cdot)$: for the shortest path from $s_\text{start}$ to $s_\text{end}$.}\\[0.1cm]
  & \multicolumn{4}{l}{$\mathcal{F}^\text{sub}(\cdot)$: for specific subtour computation.}\\[0.1cm]\hline\\[-0.3cm]
1. & \multicolumn{4}{l}{\textbf{Data Input}: $\mathcal{G}'$, $\mathcal{E}'$, $s_\text{start}$ and $s_\text{end}$.}\\[0.1cm]\hline\\[-0.3cm]
2. & \multicolumn{4}{l}{$\{s_t\}_{0}^{T^\text{hdl}}\leftarrow \mathcal{F}^\text{hdl}(\mathcal{G}',s_\text{start},s_\text{start})$.}\\[0.1cm]
3. & \multicolumn{4}{l}{$\{s_t\}_{0}^T\leftarrow \{s_t\}_{0}^{T^\text{hdl}} + \mathcal{F}^\text{fsp}(\mathcal{G}',s_\text{start},s_\text{end})$.}\\[0.1cm]
4. & \multicolumn{4}{l}{$\mathcal{G}' \leftarrow \mathcal{G}'\backslash \left\{ \{(s_t,s_{t+1})\}_0^{T^\text{hdl}-1},~\{(s_{t+1},s_t)\}_0^{T^\text{hdl}-1} \right\}$, $\tau\leftarrow 0$.}\\[0.1cm]
5. & \multicolumn{4}{l}{\textbf{While} $\tau<|\{s_t\}_0^T|-1$\textbf{:}}\\[0.1cm]
6. & & \multicolumn{3}{l}{\textbf{If} $(s_\tau,s_{\tau+1})\in\mathcal{E}'$\textbf{:}}\\[0.1cm]
7. & & & \multicolumn{2}{l}{$\{s_k^\text{sub}\}_{k=\tau+1}^{T^\text{sub}} \leftarrow \mathcal{F}^\text{sub}(s_{\tau+1},s_\tau,\mathcal{G}')$.}\\[0.1cm]
8. & & & \multicolumn{2}{l}{$ \mathcal{G}' \leftarrow \mathcal{G}' \backslash \left\{ (s_\tau,s_{\tau+1}),~(s_{\tau+1},s_\tau) \right\}$.}\\[0.1cm]
9. & & & \multicolumn{2}{l}{$ \{s_t\}_{0}^T \leftarrow \{s_k\}_{0}^\tau + \{s_k^\text{sub}\}_{\tau+1}^{T^\text{sub}} + \{s_k\}_{\tau+1}^T $.}\\[0.1cm]
10. & & & \multicolumn{2}{l}{$\mathcal{G}' \leftarrow \mathcal{G}'\backslash \left\{ \{(s_k^\text{sub},s_{k+1}^\text{sub})\}_{\tau+1}^{T^\text{sub}-1},\{(s_{k+1}^\text{sub},s_k^\text{sub})\}_{\tau+1}^{T^\text{sub}-1} \right\}$.}\\[0.1cm]
11. & & \multicolumn{3}{l}{$\tau \leftarrow \tau+1$.}\\[0.1cm]
12. & \multicolumn{4}{l}{$C\leftarrow \sum_{t=0}^{T-1} c_{s_t,s_{t+1}}$.}\\[0.1cm]\hline\\[-0.3cm]
13. & \multicolumn{4}{l}{\textbf{Output}: $\{s_t\}_{0}^{T}$ and $C$.}\\[2pt]
\end{tabular} 
\label{alg_mainAlg_T1T2} 
\end{algorithm}

Several explanatory comments are made. First, $\mathcal{G}'$ denotes the Eulerian graph augmentation of the undirected graph $\mathcal{G}$. Thus, $\mathcal{G}$ is augmented in a total minimum cost manner such that afterwards every vertex has an even degree, i.e., an even number of incident edges, however, this is subject to the constraint that all interior edges shall not be eligible as augmentation candidates. The reason therefore is to enforce path planning with forward motion only for any in-field operating vehicle. Consequently, only headland edges and island headland edges are eligible. The edges replicated from $\mathcal{G}$ for this augmentation are denoted by $\mathcal{E}'$. Due to the characteristic connectivity of $\mathcal{G}$ with each vertex being connected to at most 3 vertices and aforementioned constraint, an Eulerian graph augmentation (see \cite{bondy1976graph}) can always be constructed by pairing neighbouring vertices in a cost-minimising manner. As a consequence, an overall path length minimising field coverage route for T1 with equal start end vertex is always constructed by traversing every edge at most twice. By contrast, for T2 with $s_\text{start}\neq s_\text{end}$, some edges have to be traversed three times due to the final transition to $s_\text{end}$ after completing the coverage of all edges of $\mathcal{G}$.

Second, function $\mathcal{F}^\text{hdl}(\cdot)$ returns $\{s_t\}_{0}^{T^\text{hdl}}$, which is the concatenation of the sequence of vertices tracing the headland path in counter clockwise (CCW) direction from $s_\text{start}$ to $s_\text{start}$. The CCW direction is a choice, motivated by the desire to ultimately obtain consistent circular pattern-like path planning to be detailed below. Importantly, this choice does not compromise optimality since tracing complies with $\mathcal{G}'$ and $\mathcal{G}'$ is not affected thereby. Step 3 of Algorithm \ref{alg_mainAlg_T1T2}, i.e., a shortest path computation on top of Step 2 only applies for T2 when $s_\text{start}$ differs from $s_\text{end}$. 

Third, in Step 4 all edges that were covered as part of Step 2 are removed from $\mathcal{G}'$ in both directions, whereby they are removed only once. Thus, edges stemming from the Eulerian graph augmentation as well as all interior edges and island headland  edges are unaffected and remain with $\mathcal{G}'$. Furthermore, edges that stem from the shortest path contribution due to T2 and Step 3 are not removed. The general consequence of Step 4 is  a reduction of edge candidates from $\mathcal{G}'$ that are available for future traversal in Steps 5-11.

Fourth, $\{s_t\}_0^T$ is traced throughout Steps 5-11. As soon as an edge from the Eulerian graph augmentation is traversed in Step 6, a subtour is computed in Step 7 starting from vertex $s_{\tau+1}$ and ending at $s_{\tau}$. The method for subtour computation is equivalent to a shortest path computation, however, here subject to two additional constraints plus one exploration heuristic. The two constraints are: (i) transitions along the headland path are feasible only in \emph{CCW}-direction, and (ii) only forward motion and thus no \textit{a-b-a} sequence of vertices is permitted. The exploration heuristic is crucial. It enforces exploration of any edge as soon as that edge is element of $\mathcal{E}'$ and determined to be a feasible next transition. This exploration step is necessary to avoid making part of $\mathcal{G}'$ disconnected at Step 10 of Algorithm \ref{alg_mainAlg_T1T2}, in which case full field coverage would become impossible afterwards. Without the exploration heuristic and computation of just the shortest path between  $s_{\tau+1}$ and $s_{\tau}$, the possibility of disconnection occurs in particular in case there are multiple obstacle areas.

Fifth, Steps 8-10 of Algorithm \ref{alg_mainAlg_T1T2} are responsible for removal of covered edges from $\mathcal{G}'$ and insertion of the subtour determined in Step 7 into the main sequence of vertices, $\{s_t\}_0^T$, in Step 9. Note  that the length of $\{s_t\}_0^T$ thus changes dynamically during runtime. Consequently, there are also the number of iterations according to Step 5 of Algorithm \ref{alg_mainAlg_T1T2} that change during runtime.

\begin{figure}
\centering%
\begin{tikzpicture}
\draw [black,-{Latex[scale=1.0]}] plot [rounded corners=0.25cm] coordinates { (0.35,0)(1.85,0)(1.85,2)(0.85,2)(0.85,0.1)(2.85,0.1)
};
\node[color=black] (a) at (0.85, -0.2) {$a$};
\node[color=black] (a) at (1.85, -0.2) {$b$};
\node[color=black] (a) at (1.85, 2.2) {$c$};
\node[color=black] (a) at (0.85, 2.2) {$d$};
\node[color=black] (a) at (2.85, -0.2) {$e$};
\draw [black,-{Latex[scale=1.0]}] plot [rounded corners=0.25cm] coordinates { (4,0)(6,0)(6,2)(5,2)(5,0)(8,0)(8,2)(7,2)(7,0)(8.75,0)
};
\draw [black,-{Latex[scale=1.0]}] plot [rounded corners=0.25cm] coordinates { (4.9,2)(4,2)};
\draw [black,-{Latex[scale=1.0]}] plot [rounded corners=0.25cm] coordinates { (6.9,2)(6.1,2)};
\draw [draw=green,draw opacity=0.5, line width=4pt] plot [rounded corners=0.25cm] coordinates { (4,0)(8,0)(8,2)(4,2)};
\draw [draw=green,draw opacity=0.5, line width=4pt] plot [rounded corners=0.25cm] coordinates { (4,0)(8.75,0)};
\end{tikzpicture}
\caption{(Left) Sketches of the field coverage pattern naturally resulting from the application of Algorithm \ref{alg_mainAlg_T1T2} for convex field shapes and in the absence of obstacle areas. The sequence of vertices, $\{a,b,c,d,e\}$, exemplifies the path planning for coverage of two straight edges $(a,d)$ and $(b,c)$. (Right) Concatenation of two patterns and emphasis of two illustrative path transitions.}
\label{fig_pattern}
\end{figure}
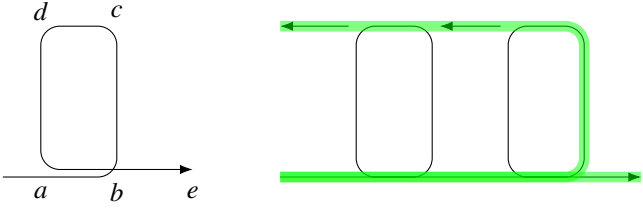

Sixth, in particular for convex field shapes and in the absence of obstacle areas, the combination of enforcing (i) traversal along all headland edges\footnote{It should be emphasised that headland edges and island headland  edges are distinguishable according to sect. \ref{subsec_preliminaries}. The CCW-direction constraint only holds for headland edges, but not for island headland  edges. } only in CCW direction, (ii) traversal of all interior edges only once, and (iii) the corresponding Eulerian graph augmentation only along headland and island headland edges causes by design a certain circular path planning pattern visualised in Fig. \ref{fig_pattern}. On the one hand, this pattern is favourable for partial field coverage as discussed in \cite{plessen2018partial}, but it is also optimal when it results from the application of Algorithm \ref{alg_mainAlg_T1T2}. This follows directly from the fact that Algorithm \ref{alg_mainAlg_T1T2} works on the full Eulerian graph $\mathcal{G}'$, which ensures a minimum cost tour. The pattern is not optimal, particularly in scenarios with multiple obstacle areas. However, then it naturally does also not occur in the solution of Step 13. Examples that illustrate this further are discussed in sect. \ref{sec_IllustrativeEx}. 

Seventh, any subtour, $\{ s_k^\text{sub}\}_{\tau+1}^{T^\text{sub}}$, is inserted into the main sequence, $\{ s_t\}_{0}^{T}$, at Step 9 of Algorithm \ref{alg_mainAlg_T1T2} immediately as soon as the subtour becomes available for traversal. As a consequence, interior edges are always covered first before any continuation along the headland path until the next cover of interior edges occurs according to a next inserted subtour. Furthermore, in case patterns according to Fig. \ref{fig_pattern} are determined from Step 7 for subtours (e.g., in the absence of obstacle areas and for convex fields), interior edges are sequentially covered in \emph{pairs} to form a concatenation of multiple of these patterns. This is favourable for the derivation of Algorithm \ref{alg_mainAlg_T3toT8} for partial field coverage.

Ultimately, the output of Algorithm \ref{alg_mainAlg_T3toT8} in Step 13 summarises the total accumulated cost (path length), $C$, computed in Step 12, and the corresponding sequence of vertices, $\{s_t\}_0^T$, of the path for full field coverage according to T1 and T2.

\subsection{Partial Field Coverage: T3 to T8\label{subsec_T3toT8}}

Algorithm \ref{alg_mainAlg_T3toT8} is proposed for partial field coverage path planning according to routing tasks T3 to T8.

\begin{algorithm}
\caption{Partial Field Coverage for T3 to T8}
\begin{tabular}{ l l l l l}
0. & \multicolumn{4}{l}{\textbf{Required Subfunctions} (brief description): }\\[0.1cm]
  & \multicolumn{4}{l}{$\mathcal{F}^\text{seq}(\cdot)$: for sequential tracing of $\{s_t\}_{0}^T$.}\\[0.1cm]
  & \multicolumn{4}{l}{$\mathcal{F}^\text{csp}(\cdot)$: for concatenated shortest paths.}\\[0.1cm]
  & \multicolumn{4}{l}{$\mathcal{F}^\text{r2n}(\cdot)$: for randomly exchanging 2 neighbours.}\\[0.1cm]\hline\\[-0.3cm]
0. & \multicolumn{4}{l}{\textbf{Hyperparameters}: $N_\mathcal{I}$, $N_{\mathcal{T}^\text{abu}}$.}\\[0.1cm]\hline\\[-0.3cm]
1. & \multicolumn{4}{l}{\textbf{Data Input}: $\mathcal{L}_\mathcal{V}$, $\mathcal{L}_\mathcal{E}$, $\mathcal{G}$, $\{s_t\}_{0}^T$, $s_\text{start}$ and $s_\text{end}$.}\\[0.1cm]\hline\\[-0.3cm]
2. & \multicolumn{4}{l}{$\mathcal{P}\leftarrow \mathcal{F}^\text{seq}( \{s_t\}_{0}^T,~ \mathcal{L}_\mathcal{V},~\mathcal{L}_\mathcal{E} )$.}\\[0.1cm]
3. & \multicolumn{4}{l}{$\mathcal{I}\leftarrow \{ 0,1,\dots,|\mathcal{P}|-1\}$.}\\[0.1cm]
4. & \multicolumn{4}{l}{$C^{\text{pfc},\star},~\{s_t^{\text{pfc},\star}\}_{0}^{T^{\text{pfc},\star}} \leftarrow \mathcal{F}^\text{csp}( \mathcal{P}(\mathcal{I}),~\mathcal{G},~s_\text{start},~s_\text{end}  )$.}\\[0.1cm]
5. & \multicolumn{4}{l}{$\mathcal{T}^\text{abu}\leftarrow \{\mathcal{I}\}$, $\mathcal{I}^\star \leftarrow \mathcal{I}$, $k\leftarrow 0$.}\\[0.1cm]
6. & \multicolumn{4}{l}{\textbf{While} $k<N_{\mathcal{I}}$\textbf{:}}\\[0.1cm]
7. & & \multicolumn{3}{l}{$\mathcal{I}\leftarrow \mathcal{F}^{\text{r2n}}(\mathcal{I}^\star)$, $j\leftarrow 0$.}\\[0.1cm]
8. & & \multicolumn{3}{l}{\textbf{While} ($\mathcal{I}\in\mathcal{T}^\text{abu}$) \textbf{and} ($j<N_{\mathcal{T}^\text{abu}}$) \textbf{:}}\\[0.1cm]
9. & & & \multicolumn{2}{l}{$\mathcal{I}\leftarrow \mathcal{F}^{\text{r2n}}(\mathcal{I})$, and $j\leftarrow j+1$.}\\[0.1cm]
10. & & \multicolumn{3}{l}{\textbf{If} $\mathcal{I}\not\in\mathcal{T}^\text{abu}$\textbf{:}}\\[0.1cm]
11. & & & \multicolumn{2}{l}{$\mathcal{T}^\text{abu} \leftarrow \{    
\mathcal{I}\} + \mathcal{T}^\text{abu}\left(0:\text{min}(|\mathcal{T}^\text{abu}|,~N_{\mathcal{T}^\text{abu}}-1)\right)$.}\\[0.1cm]
12. & & & \multicolumn{2}{l}{$C^{\text{pfc}},~\{s_t^{\text{pfc}}\}_{0}^{T^{\text{pfc}}} \leftarrow \mathcal{F}^\text{csp}( \mathcal{P}(\mathcal{I}),~\mathcal{G},~s_\text{start},~s_\text{end} )$.}\\[0.1cm]
13. & & & \multicolumn{2}{l}{\textbf{If} $C^{\text{pfc}}<C^{\text{pfc},\star}$\textbf{:}}\\[0.1cm]
14. & & & & \multicolumn{1}{l}{$C^{\text{pfc},\star},\{s_t^{\text{pfc},\star}\}_{0}^{T^{\text{pfc},\star}},\mathcal{I}^\star \leftarrow C^{\text{pfc}},\{s_t^{\text{pfc}}\}_{0}^{T^{\text{pfc}}},\mathcal{I}$.}\\[0.1cm]
15. & & \multicolumn{3}{l}{$k \leftarrow k+1$.}\\[0.1cm]\hline\\[-0.3cm]
16. & \multicolumn{4}{l}{\textbf{Output}: $C^{\text{pfc},\star}$, $\{s_t^{\text{pfc},\star}\}_{0}^{T^{\text{pfc},\star}}$, $\mathcal{I}^\star$ and $\mathcal{P}$.}\\[2pt]
\end{tabular} 
\label{alg_mainAlg_T3toT8} 
\end{algorithm}

Several explanatory comments are made. First, $\mathcal{L}_\mathcal{V}$ and $\mathcal{L}_\mathcal{E}$ denote the lists of vertices and edges to be covered according to any respective task  from T3 to T8. The coverage of specific vertices may be relevant, for example, for refilling of spraying tanks at a mobile depot waiting at a specified vertex at the field headland, or for the picking up or dropping off of, e.g., fertilising material. It is also included for generality.

Second, Step 2 of Algorithm \ref{alg_mainAlg_T3toT8} is discussed. Function $\mathcal{F}^\text{seq}(\cdot)$ traces the output of T1 for T3, T5 and T7 or the output of T2 for T4, T6 and T8, before returning the  list of edges \emph{ordered} according to this tracing. Thus, this list can be written as $\mathcal{P}= \{ \mathcal{P}_0, \mathcal{P}_1, \dots, \mathcal{P}_{|\mathcal{P}|-1} \}$, whereby each element
\begin{equation}
\mathcal{P}_i = \left( \{v_i^\text{in}\},~ v_i^\text{out} \right),\quad \forall i=0,\dots,|\mathcal{P}|-1,\label{eq_def_mPi}
\end{equation}
defines a directed edge with unique vertex $v_i^\text{out}$ for both $\mathcal{L}_\mathcal{V}$ and $\mathcal{L}_\mathcal{E}$, but unique $v_i^\text{in}$ only for $\mathcal{L}_\mathcal{E}$. Thus, $(v_i^\text{in},v_i^\text{out})\in \mathcal{L}_\mathcal{E}$ and $v_i^\text{out}\in \mathcal{L}_\mathcal{V}$. The notation $\{v_i^\text{in}\}$ in \eqref{eq_def_mPi} is used since when tracing the tour  $\{s_t\}_{0}^T$ in Step 2, there may be \emph{multiple} vertices immediately preceding any $v_i^\text{out}\in \mathcal{L}_\mathcal{V}$ throughout that tour. All of these vertices are stored in lists denoted by $\{v_i^\text{in}\},~\forall i=0,\dots,|\mathcal{P}|-1$. Thus, each list $\{v_i^\text{in}\}$ has length 1 for all $v_i^\text{out}\in \mathcal{L}_\mathcal{E}$, but may have multiple entries for $v_i^\text{out}\in \mathcal{L}_\mathcal{V}$ depending on the tour $\{s_t\}_{0}^T$.

Third, in Step 3 the initial \emph{ordering} of $\mathcal{P}_i$-elements is summarised by index-list $\mathcal{I}$. For reordered instances of index-list $\mathcal{I}$ throughout Steps 6-15, the corresponding reordering of $\mathcal{P}_i$-elements shall be denoted by $\mathcal{P}(\mathcal{I})$.

Fourth, function $\mathcal{F}^\text{csp}(\cdot)$ returns a sequence of multiple concatenated shortest paths, $\{s_t^{\text{pfc},\star}\}_{0}^{T^{\text{pfc},\star}}$, and the corresponding accumulated cost, $C^{\text{pfc},\star}$. Adding vertices, $s_\text{start}$ and $s_\text{end}$, the following list can first be written,
\begin{equation}
\left\{s_\text{start}, ~\{v_{\mathcal{I}_0}^\text{in}\},~ v_{\mathcal{I}_0}^\text{out},~\{v_{\mathcal{I}_1}^\text{in}\},~ v_{\mathcal{I}_1}^\text{out},\dots,~\{v_{\mathcal{I}_{|\mathcal{P}|-1}}^\text{in}\},~ v_{\mathcal{I}_{|\mathcal{P}|-1}}^\text{out}, s_\text{end} \right\},\label{eq_def_pairwiseSPs}
\end{equation}
before \emph{pairwise} shortest paths are computed, i.e., between pairs $s_\text{start}$ and $\{v_{\mathcal{I}_0}^\text{in}\}$, between $v_{\mathcal{I}_0}^\text{out}$ and $\{v_{\mathcal{I}_1}^\text{in}\}$, and so forth, until between $v_{\mathcal{I}_{|\mathcal{P}|-1}}^\text{out}$ and $s_\text{end}$. For the case that any $\{v_{\mathcal{I}_j}^\text{in}\}$ comprises more than one vertex, the shortest path between $v_{\mathcal{I}_{j-1}}^\text{out}$ and \emph{all} of their vertices is computed. The best vertex from $\{v_{\mathcal{I}_j}^\text{in}\}$ is then selected such that the path length from $v_{\mathcal{I}_{j-1}}^\text{out}$ to it \emph{plus} the edge path length from it to $v_{\mathcal{I}_{j}}^\text{out}$ is shortest. The return values, $\{s_t^{\text{pfc},\star}\}_{0}^{T^{\text{pfc},\star}}$ and $C^{\text{pfc},\star}$, represent the vertex sequence resulting from the concatenation of all pairwise shortest paths and the corresponding accumulated path length, whereby any edges, $(\{v_{\mathcal{I}_j}^\text{in}\},~v_{\mathcal{I}_{j}}^\text{out})$, that link the different pairwise shortest paths are also included. The shortest path computation on $\mathcal{G}$ is based on \cite{dijkstra1959note}, however, here accounting for two additional constraints: (i) transitions along the headland path are feasible only in CCW-direction in accordance with the method from sect. \ref{subsec_T1andT2} for full field coverage, (ii) directed edges $(v_{\mathcal{I}_{j}}^\text{out},~\{v_{\mathcal{I}_j}^\text{in}\}),~\forall j=0,\dots,|\mathcal{P}|-1$ are prohibited from being on any corresponding shortest path between $v_{\mathcal{I}_{j-1}}^\text{out}$ and $\{v_{\mathcal{I}_j}^\text{in}\}$. The latter is done to enforce paths with forward motion only. Because of the special modeling technique with $(v_{\mathcal{I}_{j}}^\text{out},~\{v_{\mathcal{I}_j}^\text{in}\}),~\forall j=0,\dots,|\mathcal{P}|-1$, plus the two vertices $s_\text{start}$ and $s_\text{end}$, the length of \eqref{eq_def_pairwiseSPs} will always be even such that pairwise shortest path computations are always exactly possible.

Fifth, in Step 5 of Algorithm \ref{alg_mainAlg_T3toT8} a \emph{tabu list}, $\mathcal{T}^\text{abu}$, is initialised with indexing list $\mathcal{I}$, which indicates the sequence of edges and nodes to be covered according to Step 2 and 3. The currently best indexing list, $\mathcal{I}^\star$, is also initialised, before improvement iterations start from Step 6 on.

Sixth, the fundamental idea of Steps 6-15 is to iterate over indexing list $\mathcal{I}$ with the purpose of improving cost  $C^{\text{pfc},\star}$ and the corresponding sequence of vertices, $\{s_t^{\text{pfc},\star}\}_{0}^{T^{\text{pfc},\star}}$. Function $\mathcal{F}^\text{r2n}(\cdot)$ in Steps 7 and 9 randomly exchanges 2 neighbouring indices in $\mathcal{I}^\star$ and $\mathcal{I}$, respectively, to produce a new candidate list $\mathcal{I}$. It was found that in Step 7 attempting  to exchange $\mathcal{I}^\star$, instead of the last $\mathcal{I}$, improved performance. Similarly, it was found that incrementally exchanging only two neighbouring indices yielded faster solve times in contrast to randomly reshuffling the entire $\mathcal{I}$-list at every iteration. Most importantly, it was found that the employment of a tabu list, $\mathcal{T}^\text{abu}$, significantly helped increasing the likelihood and speed of finding the global optimal $C^{\text{pfc},\star}$. This is since the effect of $\mathcal{T}^\text{abu}$ and Steps 8-10 is that exploration of different $\mathcal{I}$-candidates is enforced. According to Step 11, an $\mathcal{I}$ not yet in $\mathcal{T}^\text{abu}$ is added at index 0 and all remaining elements of $\mathcal{T}^\text{abu}$ are shifted by 1 index,  thereby deleting its previous last element if necessary to ensure a finite maximum length of the tabu list. More aspects of the size of $\mathcal{T}^\text{abu}$ are discussed at the end of this section and in sect. \ref{sec_discussion}.

Seventh, the edges to be covered as a part of a routing task may be defined as \emph{undirected} edges in $\mathcal{L}_\mathcal{E}$ when input to Algorithm \ref{alg_mainAlg_T3toT8} as part of Step 1. However, after the tracing in Step 2, all edges of $\mathcal{L}_\mathcal{E}$ are automatically directed according to the transitions from $\{s_t\}_{0}^T$. Since $\mathcal{P}$ is not changed beyond Step 2 in Algorithm \ref{alg_mainAlg_T3toT8}, also the direction of edge traversals is not further changed. This is done on purpose. In combination with the method of concatenating shortest paths subject to 2 constraints outlined above, it is thereby ensured that all transitions between any headland edge and any interior edge are unique. This is favourable with respect to minimisation of the soil compacted area. If unique transitions were not enforced, then depending on a routing mission, an unconstrained shortest path computation may generate a \emph{new} transition between a headland and interior edge, which in practice due to limited turning radii of in-field operating tractors would cause newly compacted area for this transition. Consequently, the harvestable area would be destroyed and the crop lost. Furthermore, in case there are many different partial field coverage missions, the consequence may even be that at \emph{every} transition from headland to interior and vice versa there are tractor tracks in every which direction, which would be the worst-case scenario with respect to minimisation of the soil compacted area.

Eighth, the motivation for the general methodology in Algorithm \ref{alg_mainAlg_T3toT8} is underlined. In general, the partial field coverage problems can be considered as \emph{travelling salesman problems} (TSPs) subject to additional constraints. For general TSPs, the complexity increases extremely quickly with problem size. For $n$ general entities to be visited, the number of different orders in which these can be visited is $n!$. For $n=5$ this is $n!=120$, however, for $n=10$ it is already $n!=3628800$. Key notion and the main argument for the design of Algorithm \ref{alg_mainAlg_T3toT8} is that full field coverage can be considered as a special case of partial field coverage. For that particular case, the optimal solution is immediately recovered from Step 2-4 due to the fact that Algorithm \ref{alg_mainAlg_T3toT8} starts for all partial field coverage solutions always from the full field coverage solution according to Algorithm \ref{alg_mainAlg_T1T2}. By contrast, 
for alternative TSP-solution methods that would start from a more general setting, no guarantee can be provided about the retrieval of the desired original optimal full field coverage plan. Similarly, Algorithm \ref{alg_mainAlg_T3toT8} is well suited for partial field coverage applications where groups of neighbouring interior edges (pairs), or specific regions of the field are meant to be covered. This again follows from the initial tracing of the solution for full field coverage according to Step 2. Nevertheless, iteration steps 6-15, and in particular also the usage of the tabu list for exploration, are still required for generality of partial field coverage missions, for example, when multiple vertices and edges that are very distant apart need to be covered.

Ultimately, the 2 hyperparameters in Algorithm \ref{alg_mainAlg_T3toT8}, $N_\mathcal{I}$ and $N_{\mathcal{T}^\text{abu}}$, are discussed. An upper bound is $N_{\mathcal{T}^\text{abu}}\leq \left(|\mathcal{L}_\mathcal{V}| + |\mathcal{L}_\mathcal{E}|\right)!$. There is no gain from a larger tabu list since in that case $\mathcal{T}^\text{abu}$ already can accomodate all possible sequences of vertices and edges of $\mathcal{L}_\mathcal{V}$ and $\mathcal{L}_\mathcal{E}$. Furthermore, it is sensible to bound $N_\mathcal{I}\geq N_{\mathcal{T}^\text{abu}}$ since the tabu list is otherwise never filled completely. For $N_\mathcal{I}> N_{\mathcal{T}^\text{abu}}$, there is a likelihood that throughout iteration Steps 6-15 the same $\mathcal{I}$ is added multiple times to the tabu list, which does not yield exploration progress. Therefore, it is proposed that
\begin{equation}
N_\mathcal{I} = N_{\mathcal{T}^\text{abu}}.\label{eq_NIeqNtabu}
\end{equation}
Then, there is only 1 hyperparameter in Algorithm \ref{alg_mainAlg_T3toT8}, which is also easy to tune: the larger the better for exploration and finding of the optimal solution. In practice, $N_{\mathcal{T}^\text{abu}}$ may be limited  by a constrained or desired maximum solve time or above factorial bound for small problems. Examples of this are discussed in sect. \ref{subsec_ex4to6}.

\subsection{Special case: T9\label{subsec_T9}}

The last in-field routing task for shortest path planning between any 2 vertices of $\mathcal{G}$ may become relevant, for example, once a spraying tank is empty and it must be returned efficiently to a mobile depot waiting along the field boundary. The method for shortest path computation applied is identical to the one described in sect. \ref{subsec_T3toT8}, where multiple shortest paths are concatenated. By non-negativity of edge weights and connectivity of $\mathcal{G}$ there always exists a shortest path between any 2 vertices.

\begin{figure}
\vspace{0.2cm}
\centering
\includegraphics[width=8.5cm]{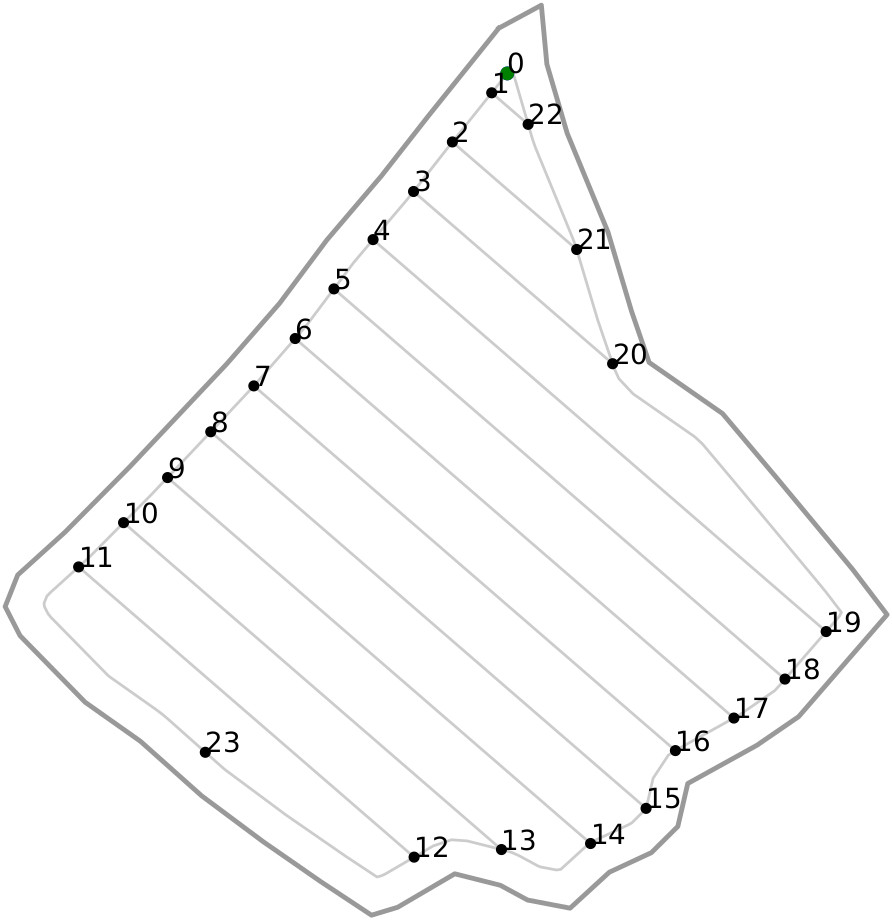} 
\caption{Field 1. A real-world field (54$^\circ$10'52.44"N,  10$^\circ$19'57.48"E) of size 13.5ha. The operating width is 36m. The vertices of $\mathcal{G}$ are labelled. The field entry and exit is vertex 0.}
\label{fig_field1}
\end{figure}

\subsection{Special case: Handling of connected subfields\label{subsec_connectedSubfields}}

As mentioned in sect. \ref{subsec_preliminaries}, \emph{separate} transition graphs are defined for all partitioned subfields and the main field. Consequently, the algorithms for T1 and T2 from sect. \ref{subsec_T1andT2} for full field coverage, for T3 to T8 for partial field coverage, and T9 as a special case are all also applied \emph{separately} to each of the subfields and the main field. However, for synchronisation 1 modification is implemented. While the headland traversal direction of the main field is defined as CCW, it is now defined as \emph{CW} for all subfields. This is the only difference and it permits insertion of subfield solutions into the main field solution sequence of vertices, while ensuring consistent travel direction along the headland path segments coinciding for both the main and all connected subfields. For visualisation, see Fig. \ref{fig_subfield}. According to above rules the travel sequence of vertices along the coinciding headland paths is $\{17,18,\dots,24\}$. An effect of this method to handle connected subfields is that part of the path coinciding for both main field and any subfield is covered four times.

\begin{figure*}
\centering
\includegraphics[width=17.5cm]{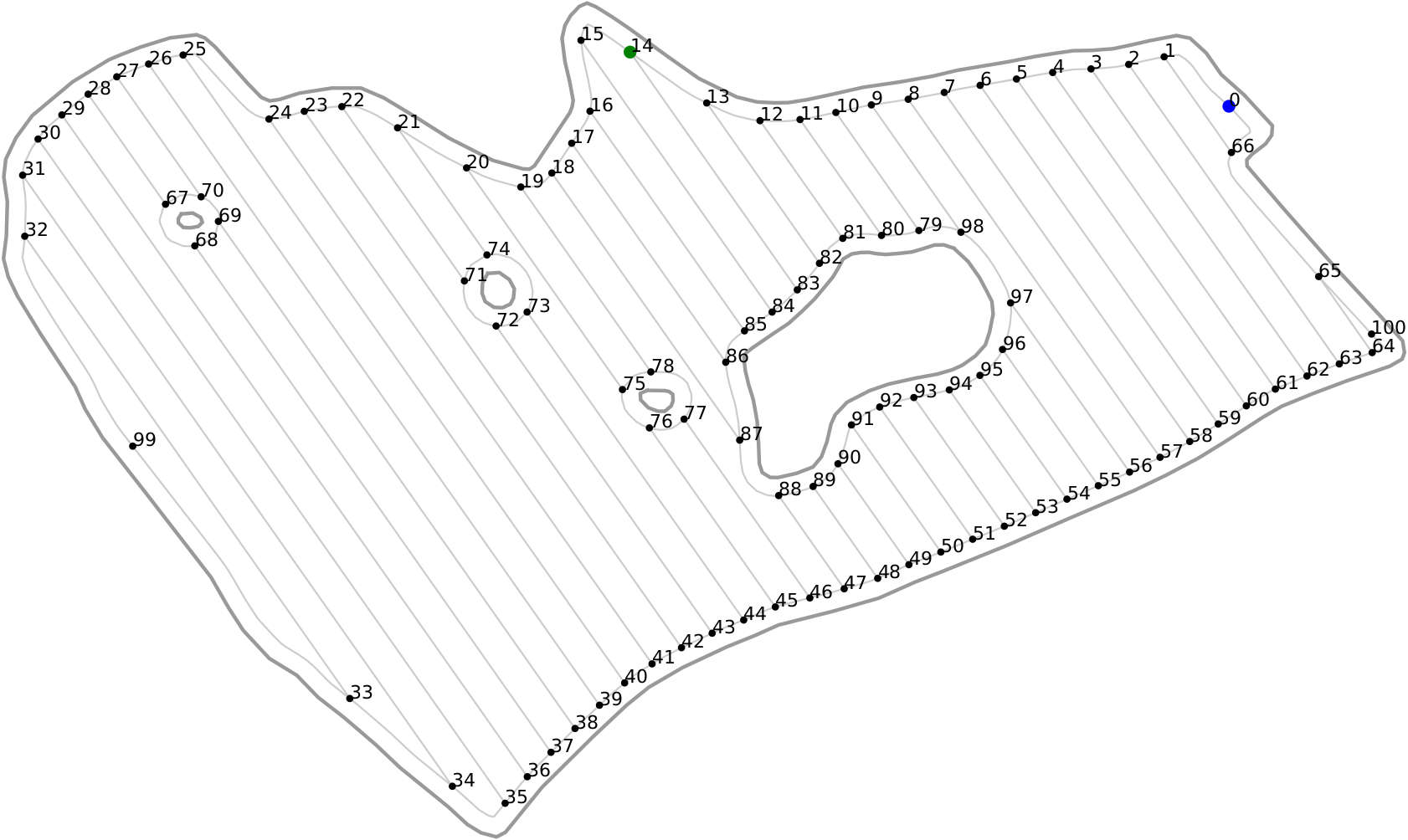} 
\caption{Field 2. A real-world field (54$^\circ$13'9.65"N, 10$^\circ$21'8.08"E) of size 74.3ha. The operating width is 36m. There are 4 obstacle areas within field contours.}
\label{fig_field2}
\vspace{0.2cm}
\end{figure*}

%
%

\section{Illustrative Examples\label{sec_IllustrativeEx}}

All methods were implemented in Python running an Intel i7-7700K CPU@4.2GHz$\times$8 processor with 15.6 GiB memory.

\subsection{Full Field Coverage Examples 1 to 3}

Algorithm \ref{alg_mainAlg_T1T2} was evaluated on three real-world examples. It should be recalled that results are guaranteed to be path length optimal since it is worked directly on the full Eulerian graph augmentation as discussed in sect. \ref{subsec_T1andT2}. Thus, results can quantitatively not be further improved. Nevertheless, path length savings, $\Delta$AB, with respect to the in practice widespread (but suboptimal) AB-pattern are stated in Table \ref{tab_field1to3} to provide a comparison. For a detailed discussion of the disadvantages of the AB-pattern, see \cite{plessen2018partial}. It is also emphasised that in \emph{addition} to optimality, Algorithm \ref{alg_mainAlg_T1T2} according to sect. \ref{subsec_T1andT2} is designed purposely (i) for compatibility with partial field coverage, and (ii) to recover the path planning pattern from Fig. \ref{fig_pattern} whenever possible. The benefits of this design are  discussed further below.

\begin{figure*}[t]
\centering
\includegraphics[width=17.5cm]{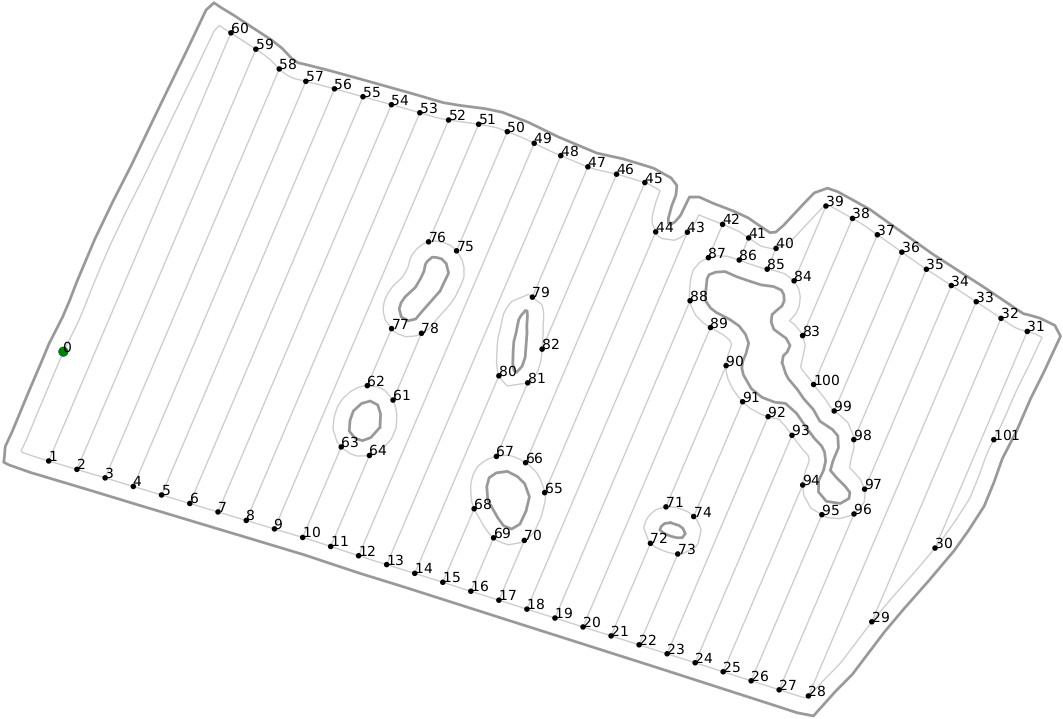} 
\caption{Field 3. A real-world field (53$^\circ$46'26.34"N,  11$^\circ$11'45.43"E) of size 62.9ha. The operating width is 36m. There are 6 obstacle areas within field contours.}
\label{fig_field3}
\vspace{0.2cm}
\end{figure*}

Example 1 is based on the field in Fig. \ref{fig_field1}. There are no obstacle areas present. The optimal sequence of vertices, $\{s_t\}_0^T$, for full field coverage according to T1 is:
\begin{align} 
&\{ 0,1,22,0,1,\textbf{2},\textbf{3},\textbf{20},\textbf{21},\textbf{2},\textbf{3},\textbf{4},5,18,19,4,5,6,7,\notag\\
&16,17,6,7,8,9,14,15,8,9,10,11,12,13,10,11,\notag\\
&23,12,13,14,15,16,17,18,19,20,21,22,0 \},
\label{nodeSequence_soln_rEx1_T1}
\end{align}
whereby the first occurence of the pattern illustrated in Fig. \ref{fig_pattern}, and which is naturally evolving from the application of Algorithm \ref{alg_mainAlg_T1T2}, is emphasised in bold. Field size, number of vertices, path length and in particular computation times for Algorithm \ref{alg_mainAlg_T1T2} are summarised in Table \ref{tab_field1to3}. When tracing \eqref{nodeSequence_soln_rEx1_T1}, there are five of the aforementioned patterns concatenated for field coverage.


Example 2 is based on the field in Fig. \ref{fig_field2}. The optimal sequence of vertices, $\{s_t\}_{0}^{T}$, for full field coverage according to T2 with $s_\text{start}=0$ and $s_\text{end}=14$ is: 
\begin{subequations}
\begin{align}
\{&0, 1, 2, 63, 64, 65, 66, 1, 2, 3, 4, 61, 62, 3, 4, 5, 6, 59, \label{nodeSeq_Field2_T2_patterns}\\
& 60, 5, 6, 7, 8, 57, 58, 7, 8, 9,10, 79, 98, \textbf{97}, \textbf{96}, \textbf{55}, \label{nodeSeq_Field2_T2_isl1apattern}\\
& \textbf{56}, \textbf{97}, \textbf{96}, \textbf{95}, 94, 53, 54, 95, 94, 93, 92, 51, 52, 93,\label{nodeSeq_Field2_T2_isl1bpattern}\\
& 92, 91, 90, 49, 50, 91, 90, 89,88, 47, 48, 89, 88, 87,\\
& 86, 17, 18, 87, 86, 85, 84, 15,16, 85, 84, 83, 82, 13,\\
& 14, 83, 82, 81, 80, 11, 12, 81, 80, 79, 98, 9, 10, 11,\label{nodeSeq_Field2_T2_endisl1}\\
&  12, 13, 14, 15, 16, 17, 18, 19, \textbf{20}, \textbf{75}, \textbf{78}, \textbf{77}, \textbf{76}, \textbf{45},  \label{nodeSeq_Field2_T2_isl1pattern}\\
& \textbf{46}, \textbf{77}, \textbf{76}, \textbf{75}, \textbf{78}, \textbf{19}, \textbf{20}, \textbf{21}, 22, 71, 74, 73, 72, 43, \label{nodeSeq_Field2_T2_isl2}\\
&  44, 73, 72, 71, 74, 21, 22, 23, 24, 41, 42, 23, 24, 25,\label{nodeSeq_Field2_T2_isl3}\\
& 26, 39, 40, 25, 26, 27, 28, 67, 70, 69, 68, 37, 38, 69, \label{nodeSeq_Field2_T2_isl4a}\\
& 68, 67, 70, 27, 28, 29, 30, 35, 36, 29, 30, \textbf{31}, \textbf{32}, \textbf{33},  \label{nodeSeq_Field2_T2_isl4b}\\
& \textbf{34}, \textbf{31}, \textbf{32}, \textbf{99}, 33, 34, 35, 36, 37, 38, 39, 40, 41, 42, \label{nodeSeq_Field2_T2_lastPattern}\\
& 43, 44, 45, 46, 47, 48, 49, 50, 51, 52, 53, 54, 55, 56,\\
& 57, 58, 59, 60, 61, 62, 63, 64, 100, 65, 66, 0,1,2,3,\\
&4,5,6,7,8,9,10,11,12,13,14\},\label{nodeSequ_field2_tosend}
\end{align}
\label{nodeSequ_field2}
\end{subequations}

Several comments are made. First, in \eqref{nodeSeq_Field2_T2_patterns} and \eqref{nodeSeq_Field2_T2_isl1apattern} multiple patterns can again be observed. Second, between \eqref{nodeSeq_Field2_T2_isl1apattern} and \eqref{nodeSeq_Field2_T2_endisl1} the coverage of the first and largest obstacle area with additional exploring of patterns outgoing from the island can be observed. The first pattern thereof is emphasised in bold in \eqref{nodeSeq_Field2_T2_isl1apattern} and \eqref{nodeSeq_Field2_T2_isl1bpattern}. The effect of exploring interior edges connected to the island headland  edges is an immediate result from the exploration heuristic as discussed in sect. \ref{subsec_T1andT2}. Third, in \eqref{nodeSeq_Field2_T2_isl1pattern}-\eqref{nodeSeq_Field2_T2_isl2} the sequence of vertices for the covering of the second island is in shown bold for emphasis. Notably, the same method to cover the third and fourth island can be observed in \eqref{nodeSeq_Field2_T2_isl2}-\eqref{nodeSeq_Field2_T2_isl3}, and also in \eqref{nodeSeq_Field2_T2_isl4a}-\eqref{nodeSeq_Field2_T2_isl4b}, respectively. However, it cannot be deduced from this that islands with four interior edges incident are always handled optimally as indicated. Here, this is merely a coincidence due to the Eulerian graph augmentation for the particular field in Fig. \ref{fig_field2}. Nevertheless, from these cases deterministic consistency of the output from Algorithm \ref{alg_mainAlg_T1T2} can be observed , which is desirable. Fourth, in \eqref{nodeSeq_Field2_T2_isl4b}-\eqref{nodeSeq_Field2_T2_lastPattern} the last pattern and coverage of the two last remaining interior edges is emphasised in bold. Afterwards, the sequence of vertices~\eqref{nodeSeq_Field2_T2_lastPattern}-\eqref{nodeSequ_field2_tosend} proceeds along the headland for final coverage of not-yet covered headland edges. Because of the characteristic Eulerian graph augmentation and previous pattern-like coverage of interior edges, every second headland edge from vertex 34 up until 0 has not been covered to this point. Fifth and ultimately, \eqref{nodeSequ_field2_tosend}  results from the fact that $s_\text{start}$ is different from $s_\text{end}$ for this example.


Example 3 is based on the field in Fig. \ref{fig_field3}. The optimal sequence of vertices, $\{s_t\}_{0}^{T}$, for field coverage according to T1 is:

\begin{subequations}
\begin{align}
&\{0, 1, 2, 59, 60, 1, 2, 3, 4, 57, 58, 3, 4, 5,6, 55, 56, 5,\label{nodeSequ_field3_T1_1strow}\\
&  6, 7, 8, 53, 54, 7, 8, \textbf{9}, \textbf{10}, \textbf{63}, \textbf{64}, \textbf{11}, \textbf{12}, \textbf{49}, \textbf{50}, \textbf{75},\label{nodeSequ_field3_T1_1st2isla}\\
& \textbf{76}, \textbf{77}, \textbf{78}, \textbf{61}, \textbf{62}, \textbf{63}, \textbf{64}, \textbf{61}, \textbf{62}, \textbf{77}, \textbf{78}, \textbf{75}, \textbf{76}, \textbf{51},\label{nodeSequ_field3_T1_1st2islb}\\
& \textbf{52}, \textbf{9}, \textbf{10}, \textbf{11}, 12, 13, 14, 80,81, 67, 66, 65, 70, 17,\label{nodeSequ_field3_T1_1st2islc}\\
& 18, 43, 44, 65, 70, 69, 68, 15, 16, 69, 68, 67, 66,45, \\
& 46, 82, 79, 80, 81, 82, 79, 47, 48, 13, 14, 15, 16, 17,\label{nodeSequ_field3_T1_end34isl}\\
& 18, \textbf{19}, \textbf{20}, \textbf{89}, \textbf{88}, \textbf{87}, \textbf{86}, \textbf{41}, \textbf{42}, \textbf{87}, \textbf{86}, \textbf{85}, \textbf{84}, \textbf{39},\label{nodeSequ_field3_T1_start56isl}\\
& \textbf{40}, \textbf{85}, \textbf{84}, \textbf{83}, \textbf{100}, \textbf{37}, \textbf{38}, \textbf{83}, \textbf{100}, \textbf{99}, \textbf{98}, \textbf{35}, \textbf{36}, \\
& \textbf{99}, \textbf{98}, \textbf{97}, \textbf{96}, \textbf{27}, \textbf{28}, \textbf{33}, \textbf{34}, \textbf{97}, \textbf{96}, \textbf{95}, \textbf{94}, \textbf{25}, \textbf{26},\\
& \textbf{95}, \textbf{94}, \textbf{93}, \textbf{92}, \textbf{23}, \textbf{24}, \textbf{93}, \textbf{92}, \textbf{91}, \textbf{90}, \textbf{71}, \textbf{74}, \textbf{73}, \textbf{72},\\
& \textbf{21}, \textbf{22}, \textbf{73}, \textbf{72}, \textbf{71}, \textbf{74}, \textbf{91}, \textbf{90}, \textbf{89}, \textbf{88}, \textbf{19}, \textbf{20}, \textbf{21}, 22,\label{nodeSequ_field3_T1_end56isl}\\
& 23, 24, 25, 26, 27, 28, 29, 30, 31, 32, 29, 30, 101, 31, \\
&32, 33, 34, 35, 36, 37, 38, 39, 40, 41, 42, 43, 44, 45, \\
& 46, 47, 48, 49, 50, 51, 52, 53, 54, 55, 56, 57, 58, 59,\\
&  60, 0\},
\end{align}
\label{nodeSequ_field3}
\end{subequations}

Several comments are made. First, in \eqref{nodeSequ_field3_T1_1strow} multiple of aforementioned patterns can be observed. Second, the sequence of vertices covering both the first and second obstacle area is in bold for emphasis throughout \eqref{nodeSequ_field3_T1_1st2isla}-\eqref{nodeSequ_field3_T1_1st2islc}. This sequence is not intuitive a priori. Edges are covered twice according to the Eulerian graph augmentation. Third, the coverage of the third and fourth island is described in \eqref{nodeSequ_field3_T1_1st2islc} -\eqref{nodeSequ_field3_T1_end34isl}. Fourth, the sequence of vertices covering the fifth and sixth island is in bold for emphasis in \eqref{nodeSequ_field3_T1_start56isl}-\eqref{nodeSequ_field3_T1_end56isl}. Similar to Example 2, the exploration of interior edges connected to the headland  island edges can be observed, which again is an immediate result of the exploration heuristic in Step 7 of Algorithm \ref{alg_mainAlg_T1T2}.

\begin{table}
\centering
\begin{tabular}{c|cccc|c}
\hline
\rowcolor[gray]{0.93} Ex. & Size & $|\mathcal{V}|$ & Path length & $\Delta$AB & $T_\text{solve}$  \\[1pt] 
\hline
1 & 13.5ha & 24 & 4624.5m & -829m/15$\%$ &  \textbf{0.0002}s \\
2 & 74.3ha & 101 & 26026.7m & -1621m/6$\%$ & \textbf{0.0015}s \\
3 & 62.9ha & 102 & 21974.3m & -2242m/9$\%$ & \textbf{0.0016}s \\
\hline
\end{tabular}
\caption{Full field coverage examples. Summary of results. Examples 1-3 correspond to Fields 1-3 in Fig. \ref{fig_field1}-\ref{fig_field3}. Computation runtimes for Algorithm \ref{alg_mainAlg_T1T2} are in bold for emphasis.}
\label{tab_field1to3}
\vspace{0.1cm}
\end{table}

\subsection{Partial Field Coverage Examples 4 to 6\label{subsec_ex4to6}}

Three partial field coverage examples are discussed. The first is based on Fig. \ref{fig_field1}, while the latter two are based on Fig. \ref{fig_field3}. It is stressed that Algorithm \ref{alg_mainAlg_T3toT8} is closely linked and explicitly builds on the solution from Algorithm \ref{alg_mainAlg_T1T2} for full field coverage as discussed in sect. \ref{alg_mainAlg_T3toT8}. This is enforced in order to make full and partial field coverage \emph{compatible} for minimisation of the soil compacted area from tractor tracks when accounting for limited turning radii of agricultural machinery. In absence of this compatibility, depending on the routing mission, unconstrained shortest path computation may generate new transitions between headland and interior edges, which in practice due to limited turning radii of in-field operating tractors would cause newly compacted areas for these transitions. Consequently, the harvestable area would be destroyed and crop lost. This lack of compatibility and thus danger of crop loss always occurs when partial field coverage routes are planned without accounting explicitly for full field coverage. As an aside, it should be noted that the proposed method of Algorithm \ref{alg_mainAlg_T3toT8} for partial field coverage can be built on \emph{any} provided solution for full field coverage. For example, instead of providing the path length optimal solution for full field coverage according to Algorithm \ref{alg_mainAlg_T1T2} as data input, alternatively, the path length suboptimal AB-pattern solution could be provided as input, $\{s_t\}_0^T$, in Step 1 of Algorithm \ref{alg_mainAlg_T3toT8}. As a result, partial field coverage plans compatible with the full field coverage plan according to the AB-pattern could be generated. In the interest of space, this section focuses on the main practical aspects and hyperparameter choices for Algorithm \ref{alg_mainAlg_T3toT8} for the path length optimal case.

For Example 4, the artificial problem setup comprises  $\mathcal{L}_\mathcal{E}=\{(6,17),~(9,14),~(20,21)\}$. This in-field routing task classifies as T5, arguably the most relevant class for partial field coverage. For the results in Table \ref{tab_ex_partial}, hyperparameters were set as $N_\mathcal{I}=6$ and $N_{\mathcal{T}^\text{abu}}=6$ since here $|\mathcal{L}_\mathcal{E}|!=6$ and according to \eqref{eq_NIeqNtabu}. The optimal sequence of vertices, $\{s_t^{\text{pfc},\star}\}_{0}^{T^{\text{pfc},\star}}$, is:
\begin{align}
& \{0, 1, 2, 3, 4, 5, 6, 7, 8, \textbf{9}, \textbf{14}, 15, 16, \textbf{17}, \textbf{6}, 7, 16, \notag\\ 
&17, 18, 19, \textbf{20}, \textbf{21}, 22, 0\},
\label{nodeSequ_field1_ex4}
\end{align}
where all edges involving $\mathcal{L}_\mathcal{E}$ are in bold for emphasis. Two comments are made. First, note that edges from $\mathcal{L}_\mathcal{E}$ are initially undirected. However, as a consequence of Step 2 in Algorithm \ref{alg_mainAlg_T3toT8}, they become directed in the list of $\mathcal{P}$. These directions are then maintained throughout as emphasised in \eqref{nodeSequ_field1_ex4}. Second, note how interior edge $(7,16)\not\in\mathcal{L}_\mathcal{E}$ is traversed as part of the shortest path towards $(20,21)$ after coverage of the 2 edges $(9,14)$ and $(17,6)$ that are element of $\mathcal{L}_\mathcal{E}$.

For Example 5, the artificial problem setup comprises 3 randomly and far apart selected vertices and edges, respectively. These are  $\mathcal{L}_\mathcal{V}=\{28,91,79\}$ and $\mathcal{L}_\mathcal{E}=\{(63,64),~(54,55),~(101,31)\}$. This in-field routing task thus classifies as T7. For the results in Table \ref{tab_ex_partial}, hyperparameters were set as $N_\mathcal{I}=50$ and $N_{\mathcal{T}^\text{abu}}=50$. The optimal sequence of vertices, $\{s_t^{\text{pfc},\star}\}_{0}^{T^{\text{pfc},\star}}$, is:
\begin{align}
& \{0, 1, 2, 3, 4, 5, 6, 7, 8, 9, 10, \textbf{63}, \textbf{64}, 11, 12, 13, 14, \notag\\ 
&15, 16, 17, 18, 19, 20, 21, 22, 23, 24, 25, 26, \textbf{27},\textbf{28},\notag\\
& 29, 30, \textbf{101}, \textbf{31}, 32, 33, 34, 97, 96, 95, 94, 93, \textbf{92}, \textbf{91},\notag\\
&  90, 89, 88, 87, 42, 43, 44, 45, 46, \textbf{82}, \textbf{79}, 47, 48, 49, \notag\\ 
&50, 51, 52, 53, \textbf{54}, \textbf{55}, 56, 57, 58, 59, 60, 0\}
\label{nodeSequ_field3_ex5}
\end{align}
where all edges involving $\mathcal{L}_\mathcal{V}$ and $\mathcal{L}_\mathcal{E}$ are in bold for emphasis. An additional comment is made. When tracing the output of T1, $\{s_t\}_{0}^{T}$, from \eqref{nodeSequ_field3} as part of Step 2 in Algorithm \ref{alg_mainAlg_T3toT8}, it is derived
\begin{equation}
\mathcal{P}_3 = \left( \{92,74\},~ 91 \right).
\end{equation}
This implies (i) that vertex $91$, which is element of $\mathcal{L}_\mathcal{V}$, is encountered at index $i=3$, and further (ii) that this vertex has \emph{2} candidate vertices, $92$ and $74$, immediately preceding vertex $91$ in $\{s_t\}_0^T$ of T1. This possibility was discussed in detail in sect. \ref{subsec_T3toT8}. As indicated in \eqref{nodeSequ_field3_ex5}, the final optimal transition is $(92,91)$.

For Example 6, the problem setup comprises 8 edges to imitate a precision agriculture application where only specific edges, however, spread over the entire field, must be covered. These are  $\mathcal{L}_\mathcal{E}=\{(1,60),~(2,59),~(19,88),~(20,89)$, $(27,96),~(97,34),~(28,33),
~(29,32)\}$. For the results in Table \ref{tab_ex_partial}, hyperparameters were set as $N_\mathcal{I}=350$ and $N_{\mathcal{T}^\text{abu}}=350$. The optimal accumulated path length minimising sequence of vertices, $\{s_t^{\text{pfc},\star}\}_{0}^{T^{\text{pfc},\star}}$, is:
\begin{subequations}
\begin{align}
&\{0, 1, \textbf{2}, \textbf{59}, \textbf{60}, \textbf{1}, 2, 3, 4, 5, 6, 7, 8, 9, 10, 11, 12, 13, \\
&14, 15, 16, 17, 18, 19, \textbf{20}, \textbf{89}, \textbf{88}, \textbf{19}, 20, 21, 22, 23, \\
&24, 25, 26, 27, \textbf{28}, \textbf{33}, \textbf{34}, \textbf{97}, \textbf{96}, \textbf{27}, 28, 29, 30,31, \label{nodeSequ_field3_ex6_rowc}\\ 
&\textbf{32}, \textbf{29}, 30, 31, 32, 33, 34, 35, 36, 37, 38, 39, 40, 41, \label{nodeSequ_field3_ex6_rowd}\\
&42, 43, 44, 45, 46, 47, 48, 49, 50, 51, 52, 53, 54, 55, \\
&56, 57, 58, 59, 60, 0\},
\end{align}
\label{nodeSequ_field3_ex6}
\end{subequations}
where all edges involving $\mathcal{L}_\mathcal{E}$ are in bold for emphasis. Two comments are made. First, in this example the second constraint discussed in sect. \ref{subsec_T3toT8} for the specific shortest path computation to enforce forward motion only becomes active. The result is the sequence of vertices in \eqref{nodeSequ_field3_ex6_rowc}, $\{28,29,30,31\}$, preceding the directed edge traveral $(32,29)$ in \eqref{nodeSequ_field3_ex6_rowd}. Second, notice how interior edge $(30,31)$  is covered \emph{twice} throughout \eqref{nodeSequ_field3_ex6}. The second traversal in \eqref{nodeSequ_field3_ex6_rowd} is part of the shortest path back to $s_\text{end}=0$, after coverage of the last remaining edge from $\mathcal{L}_\mathcal{E}$. For full field coverage, all interior edges were constrained to be covered only once as part of the Eulerian graph augmentation in order to ensure forward motion and to encourage circular pattern-like optimal path planning whenever applicable. However, for partial field coverage, this constraint is dropped to minimise path length and soil strain due to tractor tracks. Importantly, all transitions between interior edges and headlands are still fully compatible with the results for full field coverage, such that no new compacted areas due to limited turning radii of in-field operating vehicles are created.

Finally, hyperparameter choices and the role of the tabu list are discussed. It was observed that the inclusion of a tabu list, $\mathcal{T}^\text{abu}$, in Algorithm \ref{alg_mainAlg_T3toT8} significantly helped to retrieve the global optimal solution for partial field coverage tasks. Furthermore, by increasing the size of the tabu list, exploration is more enforced and the process of finding the optimum is significantly accelerated. For example, when reducing the maximum tabu list size to $N_{\mathcal{T}^\text{abu}}=25$ for both Examples 5 and 6, and to still retrieve the optimal solution, $N_\mathcal{I}$ had to be increased to $100$ and $650$, respectively. Then, the corresponding solve times for these scenarios were $0.3306$s and $2.7173$s, which are roughly twice as large as the results in Table \ref{tab_ex_partial}. To stress this more, when reducing $N_{\mathcal{T}^\text{abu}} $ to $10$ in Example 6, it had to be set $N_\mathcal{I}=3300$ before recovering the optimal solution, which resulted in $T_\text{solve}=14$s. To summarise, it was found that inclusion of the tabu list in Algorithm \ref{alg_mainAlg_T3toT8} is a simple yet effective method to enforce exploration and speed up solve times. The larger the tabu list the better the exploration throughout Algorithm \ref{alg_mainAlg_T3toT8}.


\begin{table}
\centering
\begin{tabular}{c|cc}
\hline
\rowcolor[gray]{0.93} Ex. & Path length & $T_\text{solve}$  \\[1pt] 
\hline
4 & 1924.5m & \textbf{0.0012}s \\
5 & 4482.9m & \textbf{0.1644}s \\
6 & 7413.4m & \textbf{1.3984}s \\
\hline
\end{tabular}
\caption{Partial field coverage examples. Summary of results. Example 4 refers to Field 1, while Example 5 and 6 correspond to Field 3. Computation runtimes for Algorithm \ref{alg_mainAlg_T3toT8} are in bold for emphasis. Hyperparameter choices used for the 3 examples are $(N_\mathcal{I},N_{\mathcal{T}^\text{abu}})=(6,6)$, $(50,50)$ and $(350,350)$, respectively.}
\label{tab_ex_partial}
\vspace{0.1cm}
\end{table}

%
%
%

\section{Benefits and Limitations\label{sec_discussion}}

One benefit of the proposed methods is very small $T_\text{solve}$, which demonstrates its computational efficiency. As Table \ref{tab_field1to3} demonstrates, this particularly holds for full field coverage tasks and is a consequence of starting with an Eulerian graph $\mathcal{G}'$ before subsequently removing covered edges such that the set of feasible edge transitions shrinks with iterations, which further accelerates runtimes. By contrast, as Table \ref{tab_ex_partial} demonstrates for partial field coverage applications, $T_\text{solve}$ is typically higher. Similar to travelling salesman problems, here with additional constraints enforcing forward motion, and traversal along the headland allowed only in CCW-direction, the sequence to trace a subset of vertices or edges is not straightforward to compute, particularly if multiple obstacle areas are present.

Another benefit of proposed methods is the total absence of hyperparameters in Algorithm \ref{alg_mainAlg_T1T2} and the presence of only two hyperparameters in Algorithm \ref{alg_mainAlg_T3toT8}; preferably just 1 according to \eqref{eq_NIeqNtabu}. As pointed out towards the end of sect. \ref{subsec_ex4to6}, the larger the maximum tabu list size, $N_{\mathcal{T}^\text{abu}}$, the better for enforcing exploration. For a very small number of vertices and edges to cover, one can select $N_{\mathcal{T}^\text{abu}}=(|\mathcal{L}_\mathcal{E}| + |\mathcal{L}_\mathcal{V}|)!$, which  guarantees that all possible sequences of vertices and edges to be covered will be tested.

The main practical limitation of proposed methods is nonintuitive path planning, in particular in presence of multiple obstacle areas. One may argue that even if the returned sequence of vertices is path length optimal, it may still be not implementable. As long as tractors are not fully automatically following path plans, for example, similar to the method of \cite{plessen2017reference}, fully optimised path plans may not be practical. This is because the vehicle driver has to be ``glued to'' a navigation screen and audio commands to follow an nonintuitive path plan while simultaneously concentrating on keeping track, which may be stressful to the driver. However, it is emphasised that this aspect is explicitly addressed and mitigated by the design of Algorithm \ref{alg_mainAlg_T1T2} and \ref{alg_mainAlg_T3toT8} through the enforcement of pattern-like field coverage whenever applicable, which (i) maintains optimality since being based on $\mathcal{G}'$, and which (ii) can favourably  be translated to consistent rule-based driving instructions at least for all convex fields, such that (iii)  nonintuitive paths remain only for field portions with multiple obstacle areas.

\section{Conclusions\label{sec_conclusion}}

This paper discussed optimal in-field routing for full and partial field coverage with arbitrary non-convex fields and multiple obstacle areas. It is distinguished between nine different in-field routing tasks: two for full field coverage, seven for partial field coverage and one for shortest path planning between any two vertices of the transition graph. There was differentiation between equal and different start and end vertices for a task, coverage of all edges (full field coverage), or coverage of only a subset of vertices, and a subset of edges or combinations (partial field coverage). A key notion is how to efficiently combine the coverage of headland and island headland edges together in combination with the coverage of all interior edges. Starting from an Eulerian graph augmentation, proposed algorithms encourage a particular circular-like pattern whenever applicable without compromising optimality, such that field coverage is consistent for convexly shaped fields. For arbitrary non-convex fields and with multiple obstacle areas, the resulting path guidance is not any more intuitive, however the path length is optimal. Proposed methods are primarily developed for spraying and fertilising applications with larger working widths for in-field operating vehicles. The handling of subfields connected to the main field was discussed. The proposed solution for partial field coverage starts from the solution for full field coverage to consistently comply with transitions between interior and headland edges by accounting for limited turning radii of agricultural vehicles, and thereby ensuring that no new tractor tracks are generated in view of compacted area minimisation. For partial field coverage, the benefit of employing a tabu list in the solution algorithm for improved exploration was highlighted. Proposed methods were illustrated by means of six experiments on three real-world fields, with a focus on demonstrating low computation runtimes and the explicit mentioning of sequences of vertices to emphasise aspects of the presented path planning methods.

\bibliographystyle{model5-names} 
\bibliography{mybibfile.bib}
\nocite{*}







\end{document}